\documentclass[preprint,showpacs,preprintnumbers,amsmath,amssymb]{revtex4}
\usepackage{graphicx}
\usepackage{dcolumn}
\usepackage{bm}
\usepackage{epsfig}

\begin{document}

\title{Quantum mechanics of the two-dimensional circular billiard 
plus baffle system and half-integral angular momentum}

\author{R. W. Robinett} \email{rick@phys.psu.edu}
\affiliation{%
Department of Physics\\
The Pennsylvania State University\\
University Park, PA 16802 USA \\
}%

\date{\today}

\begin{abstract}
We examine the quantum mechanical eigensolutions of the two-dimensional
infinite well or quantum billiard system consisting of a circular
boundary with an infinite barrier or baffle along a radius. Because of the
change in boundary conditions, this system includes quantized angular 
momentum values corresponding to half-integral multiples of $\hbar/2$. 
We discuss the resulting energy eigenvalue spectrum and visualize some 
of the novel energy eigenstates found in this system. We also discuss the
density of energy eigenvalues, $N(E)$, comparing this system to the
standard circular well. These two billiard geometries have the same
area (A=$\pi R^2$), but different perimeters 
($P=2\pi R$ versus $(2\pi + 2) R$), and we compare both cases to 
fits of  $N(E)$ which make use of purely geometric arguments involving
only $A$ and $P$.
We also point out connections between the angular solutions of this
system and the familiar pedagogical example of the one-dimensional infinite 
well plus $\delta$-function potential.

\end{abstract}

\pacs{03.65.Ge, 03.65.Sq}

%
%

\maketitle

\noindent
{\Large {\bf 1.~Introduction}}
\vskip 0.25cm

Two-dimensional quantum systems can offer a range of interesting 
new features compared to more familiar 1D systems, including less trivial
implementation of boundary conditions leading to energy quantization as well
as more sophisticated connections between conserved quantum numbers and
symmetries of the system. They can provide model systems to probe
the connections between classical and quantum chaos, as in billiard
geometries, and they also find  proverbial 'real-life' experimental 
realizations in a number of surface systems, such as atomic corrals.
Two-dimensional systems can also provide the first chance to utilize 
important semi-classical methods such as periodic orbit theory 
\cite{gutzwiller,brack}, where the detailed structure (specifically 
the oscillatory component) of the energy level density, $N(E)$, can be 
very directly connected to the closed classical trajectories in the system.
Finally, students may also find connections between solutions of
the 2D Schr\"odinger equation for infinite well or billiard systems and
more familiar solutions of wave equations for membranes (drumheads) from
mathematical physics courses.

Such 2D quantum infinite well or billiard systems provide practice in the
practical application of boundary conditions on the quantum wavefunction
arising from purely geometric considerations and a number of example of
two-dimensional infinite well systems have been discussed in the
literature. The 2D infinite square well \cite{liboff} can be studied
as an example of the product of two separable one-dimensional problems,
but also in the context of energy level degeneracy, either due to
obvious symmetries or for more subtle reasons \cite{accidental_degeneracy}.
The problem of a $45^{\circ}-45^{\circ}-90^{\circ}$ isosceles
triangle billiard \cite{isoceles} can be easily solved using linear 
combination solutions of the 2D square well, while the problem of 
the equilateral ($60^{\circ}-60^{\circ}-60^{\circ}$) triangle
billiard 'footprint' has been solved in closed form by a number of
authors \cite{canadian} -- \cite{robinett_annals}, using a variety of methods.

The circular infinite well, defined by the potential energy function
\begin{equation}
      V(r,\theta) = V(r)  = \left\{ \begin{array}{ll}
               0 & \mbox{for $r<R$} \\
               \infty & \mbox{for $r\geq R$}
                                \end{array}
\right.
\end{equation}
is equivalent to the problem of a classically vibrating circular membrane
\cite{mathews}, but has also been considered in detail in the quantum
mechanical case \cite{robinett_circular_well}. This  system can
provide students with their first introduction to quantized angular
momentum values in a simple context and is the starting point for our
discussion. 

In this note,  we will extend the analysis of the 'standard' circular
well (shown in Fig.~1(a)) to a similar infinite well or quantum billiard
system consisting of the same circular 'footprint', but with the
addition of an infinitely narrow and infinitely high wall along the
$\theta =0$ direction, which we will call a baffle, as shown in Fig.~1(c).
The presence of this additional constraint changes the application of
the boundary conditions, and hence the allowed quantum numbers, both the
quantized values of angular momentum, as well as the energy eigenvalue
spectrum.  In the next section, we will review the solutions of the
full circular well, as well as the related special case of the 
'half-circular' well (as shown in Fig.~1(b)), and then use these results 
to derive the
quantum eigensolutions of the circular well plus baffle case. We find
that half-integral values (in units of $\hbar$) of quantized angular 
momentum are now allowed and we also visualize some of the novel 
position-space eigenfunctions

In Sec.~3, we then calculate the distributions of energy levels, $N(E)$, 
for the full-circular well and circular well plus baffle case, comparing 
both results to predictions made using purely geometric arguments requiring 
only knowledge of the area and perimeter of the two-dimensional systems, 
finding excellent agreement with formulae familiar in the mathematical 
literature. We emphasize the importance and relative straightforwardness
of such analyses for 2D billiard systems. 
We then discuss, in Sec.~4, the similarities between 
the circle plus baffle system and the familiar pedagogical case of the 
one-dimensional infinite well plus repulsive $\delta$-function potential 
\cite{gettys} -- \cite{delta_revival}, which illustrates how the solutions
of the full circle problem are continuously related to the new half-integral 
angular momentum states. We end by presenting our conclusions and suggesting 
additional exercises of this type, involving other novel 2D billiard systems 
which can be analyzed in terms of an energy level density.

\vskip 0.35cm
\noindent
{\Large {\bf 2.~Circular well and variations}}
\vskip 0.35cm
 
We begin by reviewing the derivation of the solutions of the circular
infinite well. The two-dimensional, time-independent, free-particle 
Schr\"odinger equation, in the relevant polar coordinates, is written 
in the form
\begin{equation}
- \frac{\hbar^2}{2\mu}
\left(
\frac{\partial^2}{\partial r^2} 
+ 
\frac{1}{r} \frac{\partial}{\partial r} 
+ 
\frac{1}{r^2}\frac{\partial^2}{\partial \theta^2} 
\right)
\psi(r,\theta) = E \psi(r,\theta)
\, .
\end{equation}
For notational convenience, we have labeled the particle mass as $\mu$,
in order to avoid confusion with familiar quantum numbers.
We assume a separable solution of the form $\psi(r,\theta) = R(r)
\Theta(\theta)$, and obtain the angular and radial equations
\begin{equation}
\frac{d^2\Theta_{(m)}(\theta)}{d \theta^2} = -m^2 \Theta_{(m)}(\theta)
\label{angular_se}
\end{equation}
and
\begin{equation}
\frac{d^2R(r)}{dr^2} + \frac{1}{r}\frac{dR(r)}{dr} 
-
\frac{m^2}{r^2} R(r) = -k^2 R(r)
\end{equation}
where $k = \sqrt{2\mu E/\hbar^2}$. Initially one can think of
$m^2$ as simply a separation constant to be determined. 

The angular equation has (normalized) solutions of the form
\begin{equation}
\Theta_{(m)}(\theta) =\frac{1}{\sqrt{2\pi}} e^{im \theta}
\label{angular_eigenfunctions}
\end{equation}
and the requirement that the solutions be single-valued under 
redefinitions of the angular variable, namely
$\Theta_{(m)}(\theta + 2\pi) = \Theta_{(m)}(\theta)$, implies that
\begin{equation}
e^{im(\theta+2\pi)} = e^{im\theta}
\qquad
\Longrightarrow
\qquad
m = 0, \pm 1, \pm 2, \pm 3,...
\end{equation}
giving the familiar integral values of quantized angular momentum.
Because of the central nature of the potential, angular momentum is
also conserved, with the $\Theta_{(m)}(\theta)$ being the eigenfunctions
of $\hat{L}_{z} = (\hbar/i)(\partial/\partial \theta)$ with eigenvalues
$m\hbar$.

The resulting equation for the radial component, when written in terms
of the variable $z=kr$,  becomes,
\begin{equation}
\frac{d^2R(z)}{dz^2} + \frac{1}{z}\frac{dR(z)}{dz}
+ 
\left( 1 - \frac{m^2}{z^2}\right) R(z)
\end{equation}
which can be recognized as Bessel's (cylindrical) equation. 
This has two linearly
independent solutions for each value of $|m|$, the so-called
regular $J_{m}(z)$ (well-behaved as $z \rightarrow 0$) and the 
singular $Y_{m}(z)$ (divergent as $z \rightarrow 0$) solutions. Only the
well-behaved $J_{m}(z)$ are used here, with the energy eigenvalues
determined by the boundary condition at the circular boundary,
namely $J_{m}(kR) = 0$. The resulting energy eigenvalues are then given
by
\begin{equation}
E_{(m,n_r)} = \frac{\hbar^2 k_{(m,n_r)}^2}{2\mu}
= \frac{\hbar^2}{2\mu R^2} \left[z_{(m,n_r)}\right]^2
\end{equation}
where $z_{(m,n_r)}$ is the $n_r$-th zero of the regular Bessel
function $J_{m}(z)$. The number of radial nodes is then given by
$n_r-1$. The properly normalized radial wavefunctions are given by
\begin{equation}
{\cal J}_{(m,n_r)}(kr)
\equiv
N_{(m,n_r)} J_{(m,n_r)}(kr)
\qquad
\mbox{where}
\qquad
\left[N_{(m,n_r)}\right]^2 \int_{0}^{R} \left[J_{(m,n_r)}(kr)\right]^2\,
r\,dr =1
\end{equation}
with $k = z_{(m,n_r)}/R$.

The complete spectrum for the full circular well corresponds to one 
set of the $m=0$ solutions, but is doubly degenerate for each value of 
$|m|>0$,  since both positive and negative values of $m$ give the same 
contributions;  this corresponds physically to the equivalence of clockwise 
versus counter-clockwise motions. Some of the low-lying energy eigenvalues 
(solid lines) are illustrated in Fig.~2, ordered by angular momentum values.

The related case of the half-circular infinite well, shown in 
Fig.~1(b),  can also be solved directly using these results. The angular
eigenfunctions in Eqn.~(\ref{angular_eigenfunctions}) can also be written
in the (normalized) form
\begin{equation}
      \tilde{\Theta}_{(m)} = \left\{ \begin{array}{ll}
               1/\sqrt{2\pi}            & \mbox{for $m=0$} \\
               \cos(m\theta)/\sqrt{\pi} & \mbox{for $m=1,2,3,...$} \\
               \sin(m\theta)/\sqrt{\pi} & \mbox{for $m=1,2,3,...$}
                                \end{array}
\right.
\label{bound_state}
\end{equation}
by taking appropriate linear combinations of the $\exp(im\theta)$ solutions,
with the same pattern of degeneracies obvious (namely one $m=0$
state and doubly degenerate $|m| \neq 0$ states.) 

The $\sin(m\theta)$ solutions will still satisfy the Schr\"odinger equation
inside the half-circular well, as well as the boundary conditions
that the wavefunction vanish along both the $\theta =0$ and $\theta = \pi$ 
directions. 
Thus, with a trivial change in normalization (an extra factor of
$\sqrt{2}$ to account for the smaller 'footprint'), the $\sin(m\theta)$ 
solutions are also appropriate for the half-circular well, with the result 
that the energy eigenvalue
spectrum for that problem consists of one copy of the integral $m>0$
energy eigenvalues in Fig.~2. The fact that the 'half-circle' state has
roughly half the states of the full circle problem will be elaborated 
upon in a detailed way in the discussion of energy eigenvalue density
in Sec.~3.

Turning now to the case of the circular well plus baffle,  as shown in
Fig.~1(c), we can also make use of the forms in Eqn.~(\ref{bound_state}),
but we must also reinterpret the required boundary condition as imposed
by the baffle. The vanishing of the angular wavefunctions along the 
$\theta = 0$ line implies that the $\cos(m\theta)$ combinations 
(and the $m=0$ state) are not allowed, while the
remaining $\sin(m\theta)$ solutions must now also vanish along the
$\theta = 2\pi$ line (the same infinite wall baffle) which implies that we
must have
\begin{equation}
\sin(2 \pi m) = 0
\qquad
\Longrightarrow
\qquad
m = \frac{1}{2}, 1, \frac{3}{2}, 2, \frac{5}{2}, ...
\end{equation}
and half-integral values of the angular momentum quantum number, $m$, are now
allowed. The integral $m=1,2,3,...$ solutions from Eqn.~(\ref{bound_state})
are still allowed, this time with the same normalization as
for the full circular well, but there is a new class of angular (and hence
radial solutions) characterized by $m = (2j+1)/2 = 1/2, 3/2, 5/2, ...$.

As we will see in detail in Sec.~4, the angular eigenfunctions
corresponding to half-integral values of $m$ correspond to the even functions 
of $\theta$ (about $\theta = 0$), so we can also write these solutions in 
the form
\begin{equation}
\tilde{\Theta}_{(m)}(\theta)  = \frac{1}{\sqrt{\pi}} \sin(m|\theta|)
\qquad
\mbox{for $m = (2j+1)/2 = 1/2, 3/2, ...$}
\end{equation}
if we choose to define them over the angular interval $(-\pi,+\pi)$.
The first four lowest lying angular solutions corresponding to
$m = 1/2, 1, 3/2$, and $2$ are shown in Figs.~3(a), (b), (c), and (d)
respectively, both over the interval $(0,2\pi)$ and $(-\pi,+\pi)$,
illustrating both the symmetry as well as the 'cusp' in 
$\tilde{\Theta}_{(m)}(\theta)$ at $\theta =0$, induced by the infinite wall,
for the $m =(2j+1)/2$ values.
Because of the additional 'angular barrier', the potential is no longer
purely central, and angular momentum is not conserved. The 
$\tilde{\Theta}_{(m)}(\theta)$ are, however, still  eigenfunctions of 
$\hat{L}_{z}^2$, with eigenvalues given by $(m\hbar)^2$; this corresponds 
classically to the fact that particles would 'rebound' from the baffle, 
reversing direction, and hence the sign of angular momentum, but keeping 
the same magnitude of $L_{z}$.

The resulting energy eigenstates, $\psi(r,\theta)$, for the integral 
$m$ values are identical to the odd-parity states of the full circular well 
(since they already vanish completely along the entire $\pm x$ axis), 
but the (normalized) wavefunctions for the half-integral $m = (2j+1)/2$ 
angular momentum states states are now given by
\begin{equation}
\psi(r,\theta) = {\cal J}_{(m=j+1/2)}(kr) \,
\tilde{\Theta}_{(2j+1)/2}(\theta)
\qquad
\mbox{with}
\qquad
k = z_{(j+1/2,n_r)}/R
\, . 
\end{equation}
We can use the well-known connection between the (cylindrical) Bessel
functions, $J_{m}(z)$,  and the spherical Bessel functions, $j_{m}(z)$
(obtained from the 3D version of the free particle Schr\"odinger equation
\cite{gasiorowicz}) to write the half-integral solutions in the form
\begin{equation}
j_{m}(z) = \sqrt{\frac{\pi}{2z}} \, J_{m+1/2}(z)
\end{equation}
Thus, half of the solutions of the 2D circular well plus baffle, those
corresponding to half-integral angular momentum,  can actually be
described by integral angular momentum solutions, but of the corresponding 
3D problem.  For example, the two lowest $m$-value spherical Bessel 
functions are given by
\begin{eqnarray}
j_{0}(z) & = & \frac{\sin(z)}{z} 
\label{spherical_bessel_0} \\
j_{1}(z) & = & \frac{\sin(z)}{z^2} - \frac{\cos(z)}{z} 
\, .
\label{spherical_bessel_1}
\end{eqnarray}
The energy eigenvalues for the $m=1/2$ case then correspond to the
zeros of the 3D $j_{0}(z)$ function, namely $z = n_r\pi$, with 
$n_r = 1,2,...$ with the result
that the $m=1/2$ energies are given by the very simple formula
\begin{equation}
E_{(m=1/2,n_r)} = \frac{\hbar^2 \pi^2 n_r^2}{2\mu R^2}
\, .
\end{equation}
We illustrate the energy eigenvalues corresponding to half-integral
values of $m$ in Fig.~2, where they nicely interpolate between the
standard integral $m$ results. The number of states for the 
circular well plus baffle case is then, in some sense, exactly the
same as that for the standard circular well, and we will in fact argue
in Sec.~4, that the half-integral angular momentum states correspond
very directly to the even-parity integral-$m$ states of the full circle,
transforming into them as the effect of the baffle is slowly added.

We then plot in Figs.~4 and 5 the position-space probability density 
($|\psi(r,\theta)|^2$) for the two lowest-lying energy eigenstates 
for the $m = 1/2$ and $m = 3/2$ states respectively, which can be
compared to the more familiar looking (drumhead-like) case of $m=1$ 
shown in Fig.~5.

For all cases with $m>1/2$, because of the fairly obvious symmetries 
of the solutions, (with $s=2m$ 'spokes' along nodal lines at
$\theta = 2\pi/s$), the expectation values of 
the position variables can be easily shown to vanish, namely,
$\langle x \rangle = \langle y \rangle = 0$ for all $n_r$ values
when $m>1/2$. For the special case of $m=1/2$, however, the expectation 
value of 
$\langle \tilde{\Theta}_{(1/2)} |\cos(\theta) | \tilde{\Theta}_{(1/2)}\rangle_{m=1/2} = -1/2$ combines with the
simple form of the radial solutions from Eqn.~(\ref{spherical_bessel_0})
to give 
$\langle {\cal J}_{(1/2)}(kr)| r |{\cal J}_{(1/2)}(kr) \rangle = R/2$ 
to yield 
$\langle x \rangle_{(m=1/2,n_r)} = -R/4$ and this asymmetry is
also obvious from Fig.~4.

\vskip 0.35cm
\noindent
{\Large {\bf 3.~Energy level density and completeness of the 
energy eigenstates}}
\vskip 0.35cm

The calculation of the number of normal modes of solutions of the 
wave equation (scalar, electromagnetic, etc.)  in 2-D or 3-D cavities 
is an important problem in a number of branches of physics, with many
discussions building on early work by Weyl. Students are perhaps most
familiar with this problem in the context of the derivation of the 
Planck formula
for blackbody radiation for a simple 3D cubical cavity which is often 
included in the modern physics curriculum, but many interesting
mathematical results exist for rather general two-dimensional shapes.

A standard reference text \cite{morse} on theoretical physics shows 
that the number of normal mode wavenumbers in the range $(k,k+dk)$ 
for a 2D shape of area $A$ and perimeter $P$ is given by
\begin{equation}
dN(k) = \left[\frac{A}{2\pi} k - \frac{P}{4\pi}\right] \, dk
\end{equation}
which upon integration gives
\begin{equation}
N(k) = \frac{A}{4\pi} k^2 - \frac{P}{4\pi} k
\end{equation}
or,  in the context of solutions of the Schr\"odinger equation of
relevance here, 
\begin{equation}
N(E) = \frac{A}{4\pi} \left(\frac{2\mu}{\hbar^2} E\right)
- \frac{P}{4\pi} \sqrt{\frac{2\mu}{\hbar^2} E}
\, .
\label{weyl_prediction}
\end{equation}
The 'experimental' energy spectrum is, of course, a discontinuous
'staircase' function of the form
\begin{equation}
N(E) = \sum_{i} \theta(E-E_i)
\end{equation}
and so the Weyl-like result of Eqn.~(\ref{weyl_prediction}) will
be an approximation to a smoothed out version of the 'data'. 
 
This type of analysis can be extended \cite{monograph} to include an
additional constant (and hence subleading) term which arises from 
the consideration of
such geometrical effects as corners, curvature, and the connectivity of
the 2D domain.  Perhaps more importantly, the oscillatory behavior of 
$N(E)$ is the central theme of periodic orbit theory \cite{gutzwiller}, 
\cite{brack} and simple pedagogical examples of this have been given 
\cite{robinett_periodic_orbit} for the square, circular well, and half 
circle 'footprints'.

The result of Eqn.~(\ref{weyl_prediction}) can be rather easily
tested on familiar 2D billiard/infinite well systems such as the
square or rectangular wells, as well as both the $45^{\circ}$ isosceles 
and $60^{\circ}$ equilateral triangles, all  because of their extremely
simple energy eigenvalue formulae (consisting of simple quadratic
powers of two integral quantum numbers.) In fact, two
studies of the equilateral triangle billiard \cite{canadian},
\cite{berry} made use of this relation as a cross-check on the
completeness of the energy eigenstates they derived, confirming
that their calculated energy spectrum saturated the Weyl-like
prediction for $N(E)$.

For the cases we consider here, we wish to compare the theoretically
predicted spectrum for $N(E)$ against the geometric 'fit' of
Eqn.~(\ref{weyl_prediction}). From the differing  'footprints'
for the geometries we have considered, the relevant area ($A$)
and perimeter ($P$) values are given by 
\begin{center}
\begin{tabular}{|l|l|r|} \hline 
shape &  $A$ & $P$ \\ \hline 
full circular well & $\pi R^2$ & $2\pi R$ \\ \hline
half circular well & $\pi R^2/2$ & $(2+\pi)R$ \\ \hline 
well plus baffle   & $\pi R^2$  & $(2\pi + 2)R$  \\ \hline
\end{tabular}
\end{center}
so that the full circular well and the well plus baffle have the same
area, but differing perimeters (due to the 'intrusive' baffle.) 
Note that the effective perimeter for the well plus baffle case is
$2R$ larger than $2\pi R$ due to the fact that both the 'top' and the
'bottom' of the baffle are inside the well, each adding a factor of
$R$.

We can easily collect (using numerical calculations of the required
zeros of various order Bessel functions) large numbers of the low-lying
energy eigenvalues for the cases we consider and we plot the
resulting $N(E)$ versus $E$ using the first hundred or so lowest $E$
values for the three cases above in Fig.~7(a), as the three 'staircase'
functions shown there. The smooth curves are the predictions of
Eqn.~(\ref{weyl_prediction}), while the two dotted curves are the
predictions using only the area ($A$) terms which are clearly not a 
very good fit by themselves.  The correspondence to 
the observed energy spectra is extremely good, and the clear difference
between the full circle and half circle cases is obvious. A smaller region
(shown as a dashed box) is enlarged on the right in Fig.~7(b) and
there one can here easily see the nice distinction between the full circle
case (upper data and dashed curve) and the circle plus baffle (lower
data and solid curve) indicating that the effective values of $A$
and $P$ used above are correct.  This case is especially interesting
since it's one of the few for which the 'footprint' area is naturally
the same, while the perimeter is different. The case of the 2D infinite
square well ($L \times L$) or an equal area rectangular well 
($fL \times L/f$) with different perimeters is another such simple
example. 

We wish to especially emphasize how this type of analysis can be rather 
easily applied to a wide variety of 2D billiard systems, perhaps with 
an eye towards introducing students to numerical methods, and eventually 
to the interesting topic of periodic orbit theory. We also note that the
discussion can be made more quantitative by applying least square fits 
to the $N(E)$ versus $E$ data used in Fig.~7, using a functional form
$N(E) = aE +  b\sqrt{E}$,  to obtain numerical values to compare directly
to the results of Eqn.~(\ref{weyl_prediction}), with good agreement being
obtained using only a few hundred low-lying states.

\vskip 0.35cm
\noindent
{\Large {\bf 4.~Relationship to 1D infinite well plus $\delta$ function
problem}}
\vskip 0.35cm

In order to better understand the structure of the half-integral
angular wavefunctions derived above, and to make connection with a
familiar one-dimensional quantum system, we wish to model the effect
of introducing the baffle wall and continuously increasing its height.
This approach is, then, very similar to the often discussed 
\cite{gettys} -- \cite{delta_revival} problem of a 1D infinite square
well with a repulsive $\delta$-function potential placed at the center.
We will briefly review the methodology and results of that problem, and
then apply it to case of the circular baffle.

As an example of how the additional of a singular potential of
arbitrary 'strength' can change the energy level structure of a simple
system, we consider a symmetric infinite well potential  with walls at
$(-L,+L)$, which has even- and odd-parity energy eigenfunctions and 
eigenvalues given by
\begin{eqnarray}
\psi_{(n)}^{(+)}(x) = \frac{1}{\sqrt{L}} \cos \left( \frac{(2n-1)\pi x}{2L}\right)
& \quad \mbox{with} \quad & E_{(n)}^{(+)} = \frac{\hbar^2 \pi^2 (2n-1)^2}{8\mu L^2} \\
\psi_{(n)}^{(-)}(x) = \frac{1}{\sqrt{L}} \sin \left( \frac{(2n) \pi x}{2L} \right)
& \quad \mbox{with} \quad & E_{(n)}^{(-)} = \frac{\hbar^2 \pi^2 (2n)^2}{8\mu L^2} 
\end{eqnarray}
where once again we denote the particle mass by $\mu$.

We then introduce a singular, repulsive potential of the form 
$V(x) = \lambda \delta(x)$ to the center of the well and ask how the
energy eigenvalues and eigenfunctions are changed. The odd parity
$\psi_{(n)}^{(-)}(x)$,  which have a node at $x=0$, are unaffected by the
addition of $V(x)$, whatever the strength of the singularity, $\lambda$.
The even parity solutions are changed and we can write a very general
even solution in the form
\begin{equation}
  \psi_{(\lambda)}^{(+)}(x) = \left\{ \begin{array}{ll}
              A\cos(kx) + B \sin(kx) & \mbox{for $0 \leq x \leq  +L$} \\
              A\cos(kx) - B\sin(kx)  & \mbox{for $-L\leq  x \leq 0$}
                                \end{array}
\right.
\label{even_states}
\end{equation}
The boundary conditions at either wall (where $\psi$ must vanish)
and at the singularity (where $\psi'(0)$ is discontinuous) 
are
\begin{equation}
\psi_{(\lambda)}(-L) = \psi_{(\lambda)}(+L) = 0
\qquad
\mbox{and}
\qquad
\psi'_{(\lambda)}(0_{+}) - \psi'_{(\lambda)}(0_{-}) = 
\frac{2\mu \lambda}{\hbar^2} \psi_{(\lambda)}(0)
\end{equation}
and give the relations
\begin{equation}
A\cos(kl) + B\sin(kL) = 0
\qquad
\mbox{and}
\qquad
2Bk = \frac{2\mu \lambda}{\hbar^2} A
\end{equation}
which combine to yield  the energy eigenvalue condition
\begin{equation}
\lambda \left(\frac{\mu L}{\hbar^2}\right) = - \frac{kL \cos(kL)}{\sin(kL)}
\, .
\label{square_well_condition}
\end{equation}
For the case of $\lambda =0$ (no additional $\delta$ perturbation),
the solutions are given by $kL = (2n-1)\pi/2$ and the standard
$\psi_{(n)}^{(+)}(x)$ states are reproduced. In the limit of
$\lambda \rightarrow +\infty$, however, a constant horizontal 
line of $\lambda (\mu L/\hbar^2)$ cuts the right hand side of
Eqn.~(\ref{square_well_condition}) at $kL = n\pi$, so that the
even energy eigenvalue solutions approach those of the (unchanged) 
odd parity solutions from below, namely 
$E_{(\lambda)}^{(+)}(\lambda \rightarrow \infty) \rightarrow E_{(n)}^{(-)}$
and the energy spectrum is now doubly degenerate for each $n$ value.

For the case of the baffle added to the standard circular well, we will
model the effect of continuously 'turning on' the baffle by re-writing
the angular Schr\"odinger equation in Eqn.~(\ref{angular_se}) in the
form
\begin{equation}
- \frac{d^2\Theta_{(m)}(\theta)}{d \theta^2} 
+ g\delta(\theta) = m^2 \Theta_{(m)}(\theta)
\end{equation}
with the angles defined over the symmetric interval $(-\pi,+\pi)$.
The odd parity (in $\theta$) angular $\sin(m\theta)$ solutions
of Eqn.~(\ref{bound_state}) are unaffected by the additional
repulsive $\delta$ interaction, while we can write the even solutions,
for arbitrary values of $g$, in a very similar form to that in 
Eqn.~(\ref{even_states}), namely
\begin{equation}
  \Theta_{(m)}^{(+)}(\theta) = \left\{ \begin{array}{ll}
   A \cos(m\theta) + B \sin(m\theta) & \mbox{for $0 \leq \theta \leq +\pi$} \\
   A \cos(m\theta) -  B \sin(m\theta)  & \mbox{for $-\pi \leq  \theta \leq 0$}
                                \end{array}
\right.
\label{even_angular_states}
\end{equation}
The relevant boundary conditions are now on the continuity 
of $\Theta'(\theta)$ at $\theta = \pm \pi$ and on the correct 
discontinuity at $\theta = 0$ which give
\begin{equation}
2Bm\cos(m\pi) = 2Am\sin(m\pi)
\qquad
\mbox{and}
\qquad
2mB = gA
\end{equation}
respectively. These combine to give the condition for the quantized angular
momentum quantum numbers, $m$,  as
\begin{equation}
g = \frac{2m \sin(m\pi)}{\cos(m\pi)} \equiv f(m)
\label{eigenvalue}
\end{equation}
We plot the right-hand-side of Eqn.~(\ref{eigenvalue}) in
Fig.~8 and note that solutions of this eigenvalue problem once again
correspond to the intersections of horizontal lines of constant
$g$ with the various branches of $f(m)$. For $g=0$, namely the
case of the standard circular well without the baffle, the intersections 
are at integral
values of $m$, including $m=0$, and the even and odd angular wavefunctions
are doubly degenerate (except for $m=0$) as noted above. 
For the case of $g \rightarrow
+\infty$, the intersections arise at half-integral values of $m$, so
that $m = 1/2, 3/2, ...$ and the angular solutions are no longer 
degenerate. One can, in fact, write closed-form expressions for the
(appropriately normalized) even parity angular eigenfunctions 
in the form
\begin{equation}
\tilde{\Theta}_{(m)}^{(+)}(\theta) = \frac{\cos(m |\theta| - \phi_{m})}
{\sqrt{\pi (1 + \sin(2m\pi) cos(2\phi_{m})/2m\pi)}}
\end{equation}
where $\phi_{m}$ is given by $\tan(\phi_{m}) = g/2m$ and $m$ is
given by the solutions of Eqn.~(\ref{eigenvalue}). 
This form actually gives the correct normalization for both the 
$g \rightarrow 0$ limit of integral angular momentum values 
(both $m=0$ and $m > 0$) and the $g \rightarrow +\infty$
limit of half-integral values.

\vskip 0.35cm
\noindent
{\Large {\bf 5.~Conclusions and discussion}}
\vskip 0.35cm

We have analyzed the circular well plus baffle system in detail,
making use of the Weyl-like energy eigenvalue distribution
of Eqn.~(\ref{weyl_prediction}) for two-dimensional quantum billiard
systems. The system studied here can also be approached
as the limiting case of a circular 'slice' or 'wedge' potential
\cite{robinett_slice}. A circular infinite well,  
where the angle subtended 
by the billiard is given by  $\Phi = (1+f)\pi$,  can
interpolate between the half-circle case (when $f=0$) and the well circular
well plus baffle case (when $f=1$). The boundary conditions 
($\sin(m\Phi) = 0$) imply that 
the angular momentum values must satisfy $m = n/(1+f)$ for integral $n$ and
the one can still  easily find the zeroes of $J_{(m=n/(1+f))}(z)$ using 
standard mathematical packages. Using this type of data, one can confirm 
that the energy eigenstate distribution, $N(E)$, still tracks the result 
of Eqn.~(\ref{weyl_prediction}), with continuously varying areas and 
perimeters now  given by $A_f = (1+f)\pi R^2$ and $P_f = (2 +(1+f)\pi)R$. One
can also consider $-1< f<0$ to handle such cases as the 'quarter-circle'
billiard and even smaller slices.

Another related case which can be analyzed using these methods is the annular
infinite well or billiard \cite{robinett_annular},
\cite{robinett_annular_other}, where a second 
concentric infinite barrier at $R_{in} = fR$ (with $0<f<1$) is added, 
and the particle is confined to the 
radial region $R_{in} = fR < r < R$. The angular solutions 
(with or without a baffle) 
are easily obtained as above, while the radial eigenfunctions are linear 
combinations of both cylindrical Bessel functions,
\begin{equation}
R(r) = \alpha J_{m}(kr) + \beta Y_{m}(kr)
\, ,
\end{equation}
where the divergent $Y_{m}(kr)$ is included since the singular $r=0$ point is 
explicitly excluded. Imposing the boundary condition that 
$R(R_{in} = rR)=0$ as well as vanishing at the outside boundary,
gives the eigenvalue condition
\begin{equation}
J_{m}(kr) Y_{m}(fkR) - J_{m}(fkR) Y_{m}(kR) = 0
\end{equation}
and one can once again evaluate $N(E)$ and compare it to 
Eqn.~(\ref{weyl_prediction})
using $A_f = \pi R^2 (1-f^2)$ and $P_f = 2\pi R (1+f)$ with good agreement.
Interested students can find a variety of relatively simple 2D quantum
billiard geometries for which such an analysis is fairly straightforward.

\vskip 0.35cm
\noindent
{\Large {\bf Acknowledgments}}
\vskip 0.35cm
This work was supported in part by the National Science Foundation
under Grant DUE-9950702.

\newpage

\begin{flushleft}
{\Large {\bf 
Figure Captions}}
\end{flushleft}
\vskip 0.35cm
 
\begin{itemize}
\item[Fig.\thinspace 1.]  Geometric 'footprints' for the 
(a) circular infinite well (or billiard), 
(b) the half-circular well, 
and 
(c) the circular well with baffle, studied here.
\item[Fig.\thinspace 2.]  Energy spectrum versus quantized angular
momentum values relevant for the circular, half-circular, and
circular well with baffle cases. Energy values (in units of
$\hbar^2/2 \mu R^2$) for integral values of angular momentum 
($m = 0,1,2,...$) are shown as solid, while results for half-integral 
($m = 1/2, 3/2, ...$) are shown as dashed.
The spectrum for the full circular
well consists of one copy of the $m=0$ values and two sets of the
integral $m>0$ values, while for the half-circular well, one set of
the integral $m>0$ values gives the entire spectrum. For the circular
well plus baffle case, one set of both the integral and half-integral 
values with $m>0$ constitute the entire energy spectrum.
\item[Fig.\thinspace 3.]  Angular wavefunctions, 
$\tilde{\Theta}_{(m)}(\theta)$ versus $\theta$, 
 for half-integral and integral values of quantized
angular momentum for (a) $m=1/2$, (b) $m=1$, (c) $m=3/2$, and
(d) $m=2$, over the ranges $(0,2\pi)$ and $(-\pi,+\pi)$. The 'cusps'
at $\theta =0$ for the half-integral cases, induced by the infinite
baffle along $\theta = 0$,  are evident.
\item[Fig.\thinspace 4.]  Normalized position-space probability
densities, $|\psi(r,\theta)|^2$, for the two lowest-lying $m=1/2$ 
energy eigenstates ($n_{r} =1$ on top and $n_{r} = 2$ at the bottom).
\item[Fig.\thinspace 5.]  Same as Fig.~4, but for the two lowest-lying
$m=3/2$ states.
\item[Fig.\thinspace 6.]  Same as Fig.~4, but for the two lowest-lying
$m=1$ states. These states are also solutions for the standard full
circular well. 
\item[Fig.\thinspace 7.]  Energy eigenvalue distribution, $N(E)$
versus $E$, for the full circle and circle plus baffle (top two 
curves in (a)) and half-circle case (bottom curve in (a)). The 
dashed and solid lines are the Weyl-like predictions of
Eqn.~(\ref{weyl_prediction}), while the dotted lines are the result
of only using the first 'area' term. The dashed square part of (a)
is expanded at the right in (b) to show the fine details and the clear
differences between the full circle (top, dashed line) and circle
plus baffle (bottom, solid line) cases.
\item[Fig.\thinspace 8.]  Plot of the 'eigenvalue' condition of
Eqn.~(\ref{eigenvalue}) versus $m$ for even-parity  angular wavefunctions.
 Horizontal lines of constant
$g$ cut the curve at solutions of the eigenvalue condition for
allowed values of $m$. For $g=0$ (corresponding to no baffle), the
allowed eigenvalues are $m=0,1,2,...$ as expected, while for
$g \rightarrow +\infty$, the intersections are at half-integral
values given by $m = 1/2, 3/2, ...$ corresponding to the circle plus
baffle case.
\end{itemize}


\newpage

\noindent
\hfill
\begin{figure}[hbt]
\, \hfill \,
\begin{minipage}{0.7\linewidth}
\epsfig{file=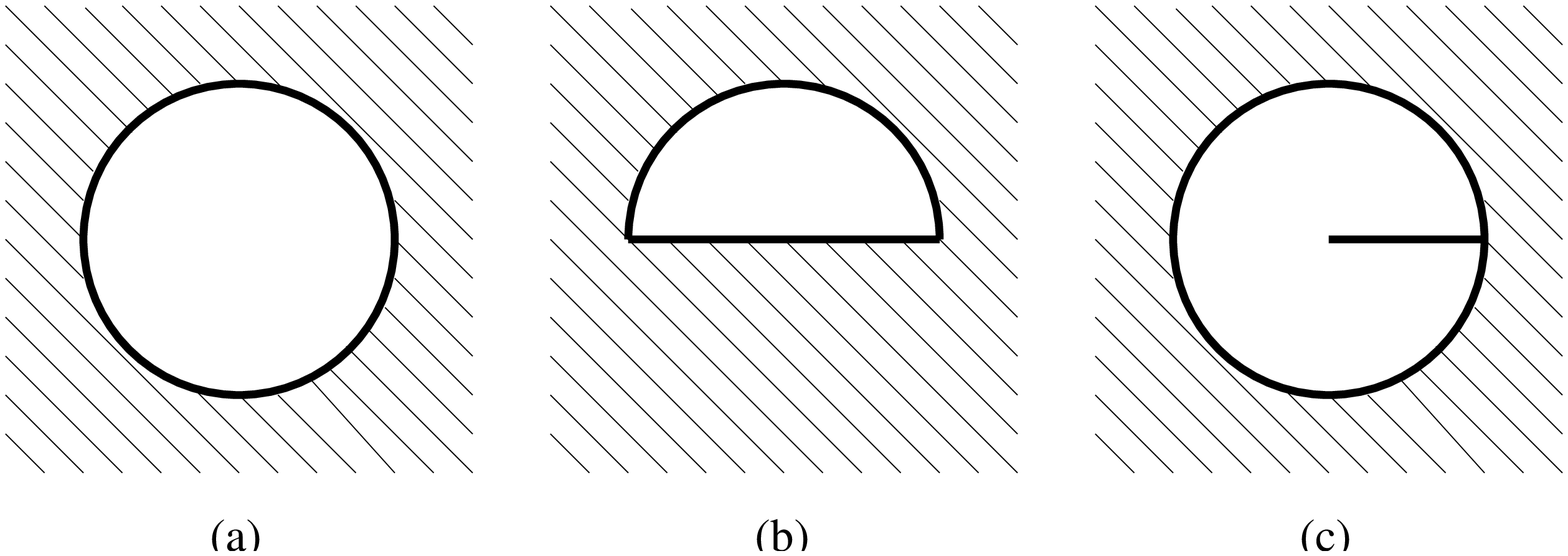,width=\linewidth}
\caption{}
\end{minipage}
\, \hfill \,
\end{figure}
\hfill

\noindent
\hfill
\begin{figure}[hbt]
\, \hfill \,
\begin{minipage}{0.7\linewidth}
\epsfig{file=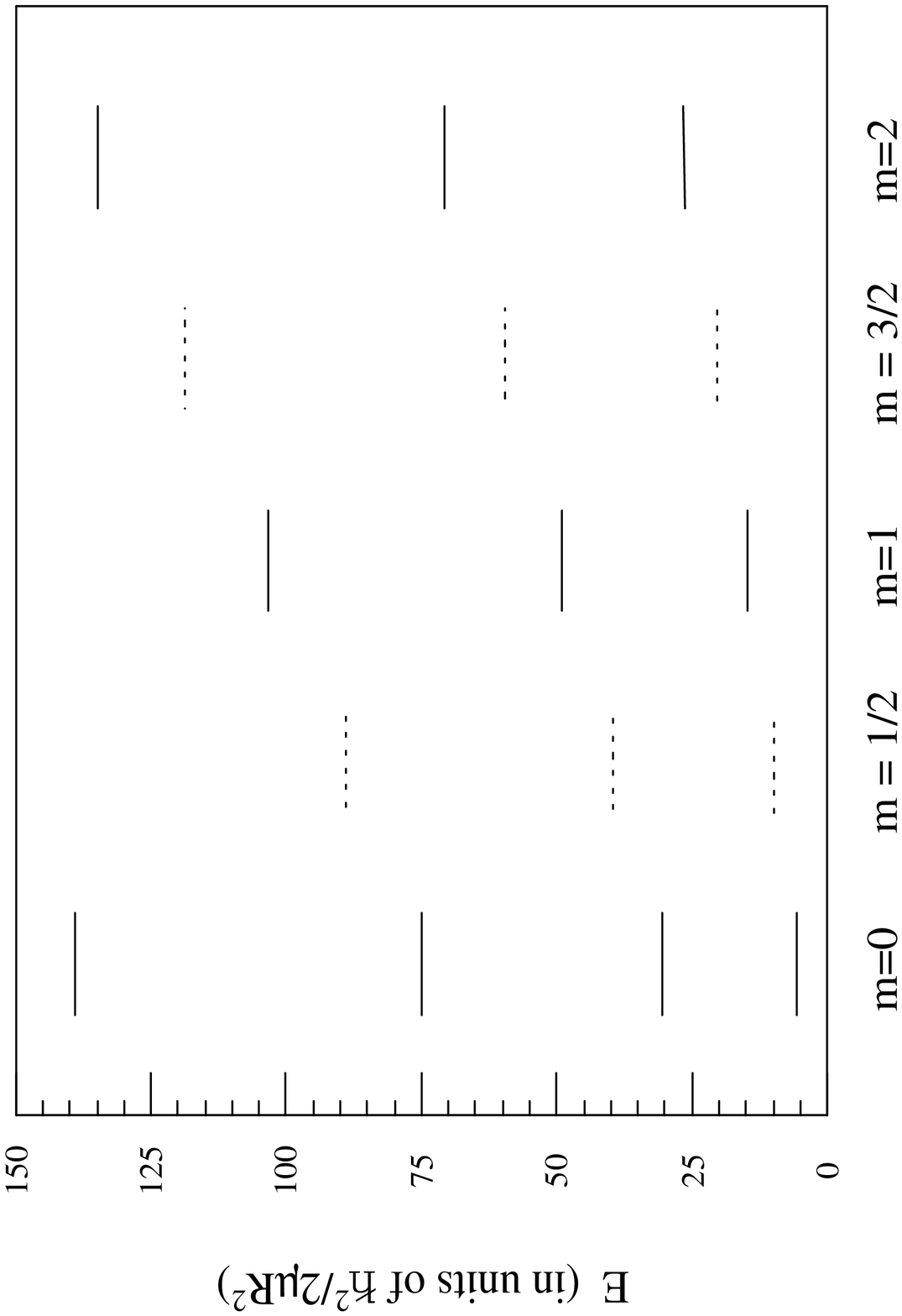,width=\linewidth}
\caption{}
\end{minipage}
\, \hfill \,
\end{figure}
\hfill

\noindent
\hfill
\begin{figure}[hbt]
\, \hfill \,
\begin{minipage}{0.7\linewidth}
\epsfig{file=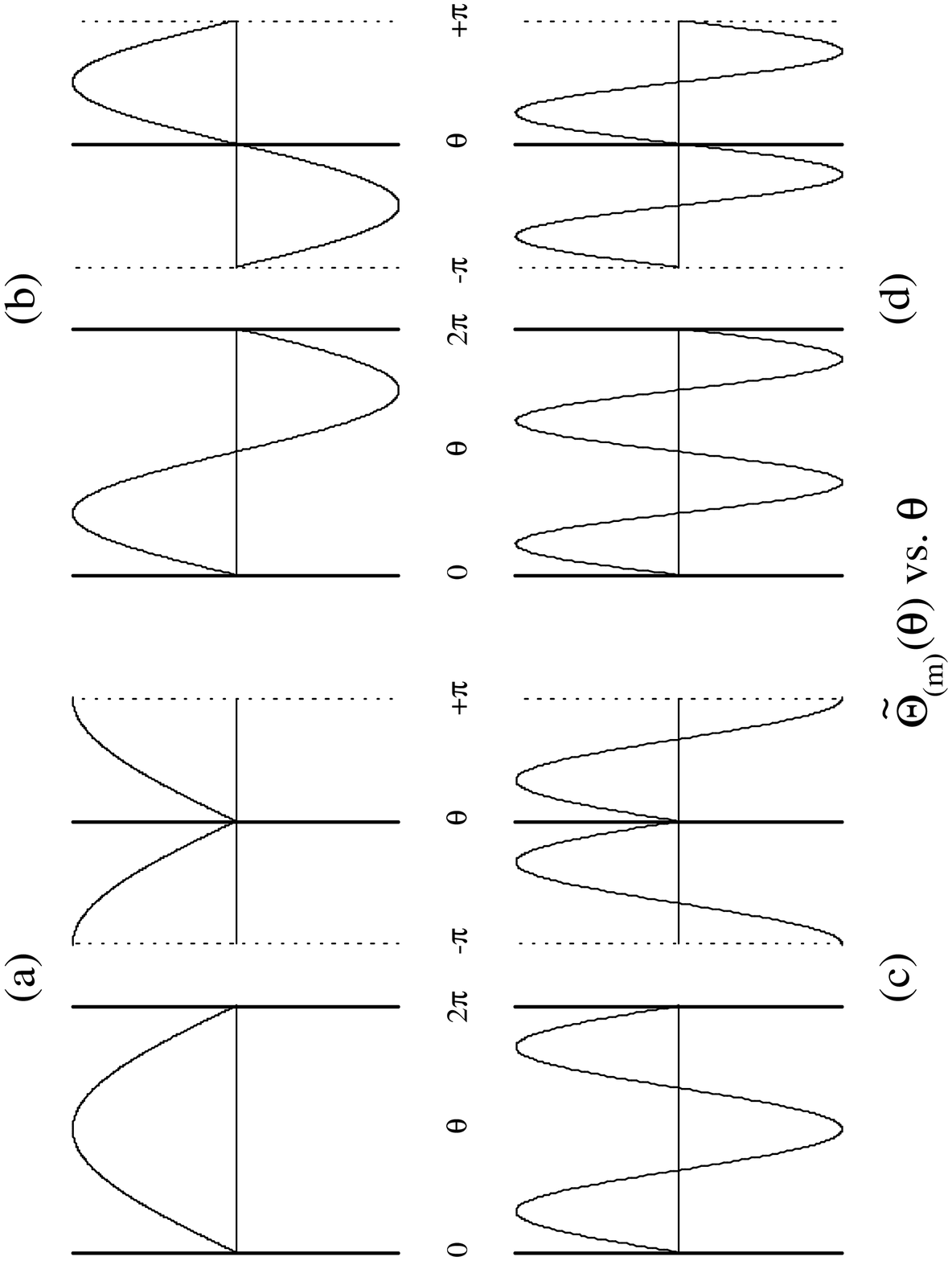,width=\linewidth}
\caption{}
\end{minipage}
\, \hfill \,
\end{figure}
\hfill

\noindent
\hfill
\begin{figure}[hbt]
\, \hfill \,
\begin{minipage}{0.8\linewidth}
\epsfig{file=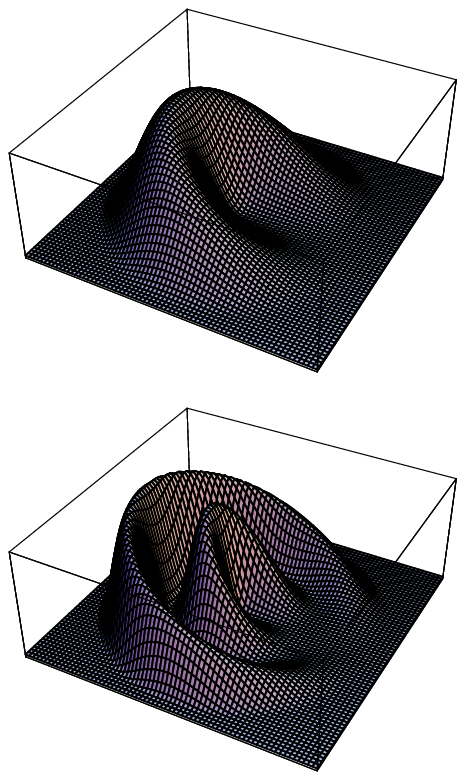,width=\linewidth}
\caption{}
\end{minipage}
\, \hfill \,
\end{figure}
\hfill

\noindent
\hfill
\begin{figure}[hbt]
\, \hfill \,
\begin{minipage}{0.8\linewidth}
\epsfig{file=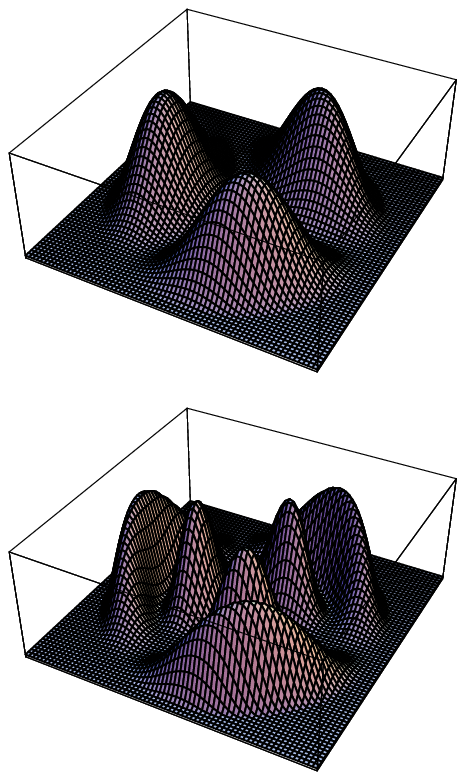,width=\linewidth}
\caption{}
\end{minipage}
\, \hfill \,
\end{figure}
\hfill

\noindent
\hfill
\begin{figure}[hbt]
\, \hfill \,
\begin{minipage}{0.7\linewidth}
\epsfig{file=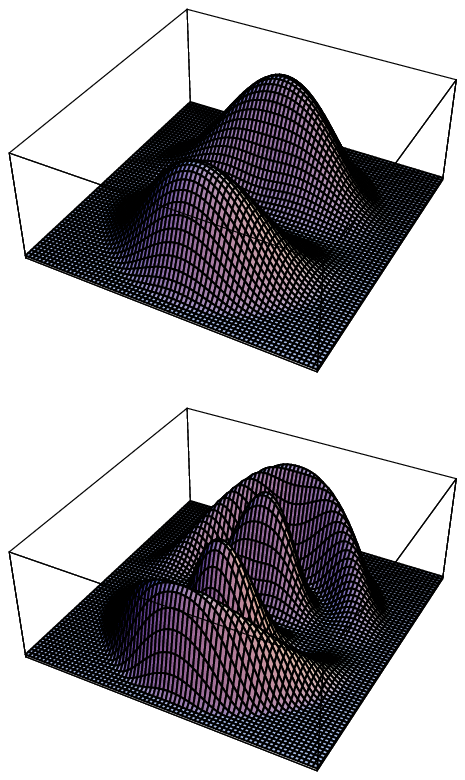,width=\linewidth}
\caption{}
\end{minipage}
\, \hfill \,
\end{figure}
\hfill

\noindent
\hfill
\begin{figure}[hbt]
\, \hfill \,
\begin{minipage}{0.5\linewidth}
\epsfig{file=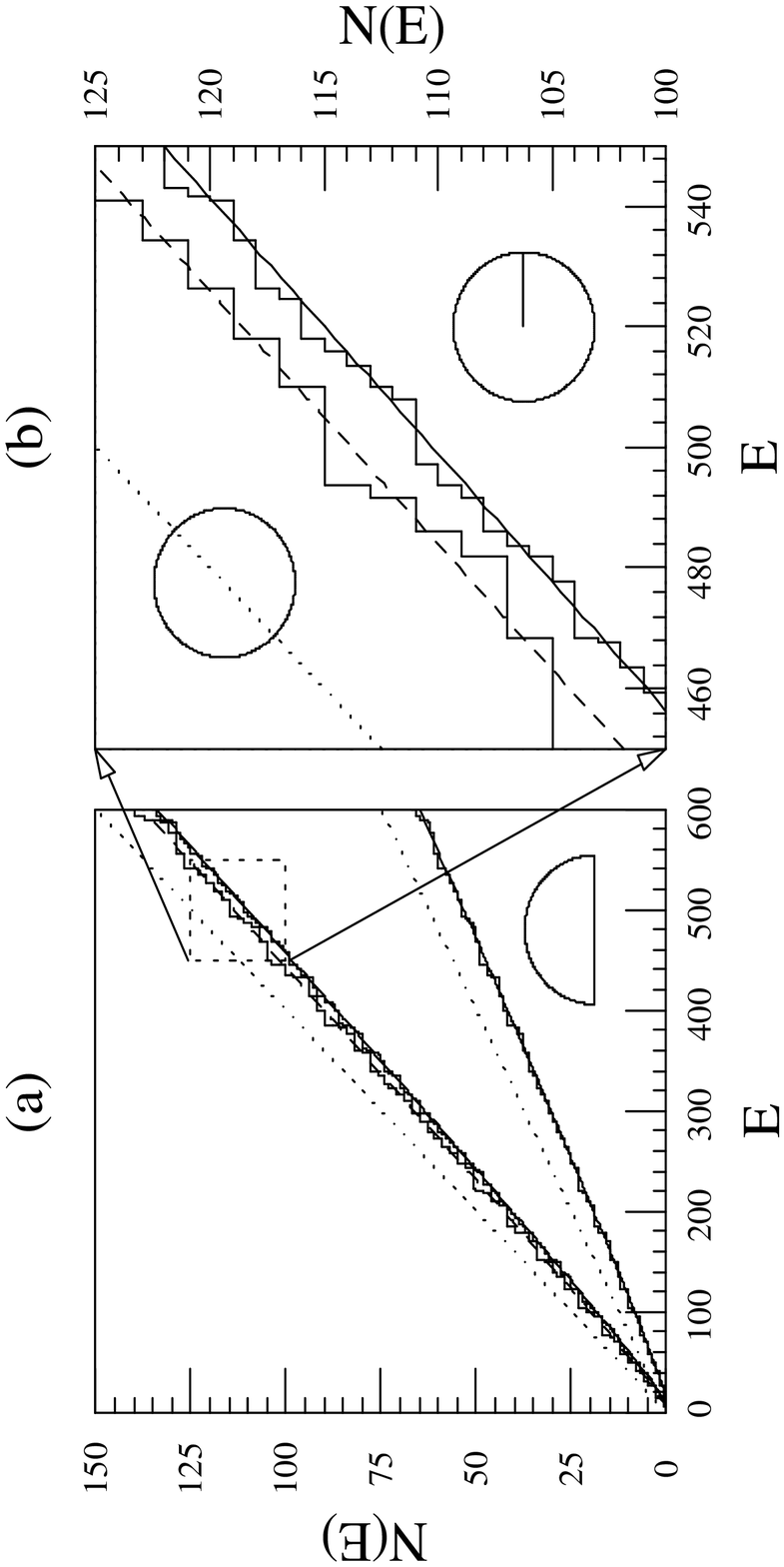,width=\linewidth}
\caption{}
\end{minipage}
\, \hfill \,
\end{figure}
\hfill

\noindent
\hfill
\begin{figure}[hbt]
\, \hfill \,
\begin{minipage}{0.5\linewidth}
\epsfig{file=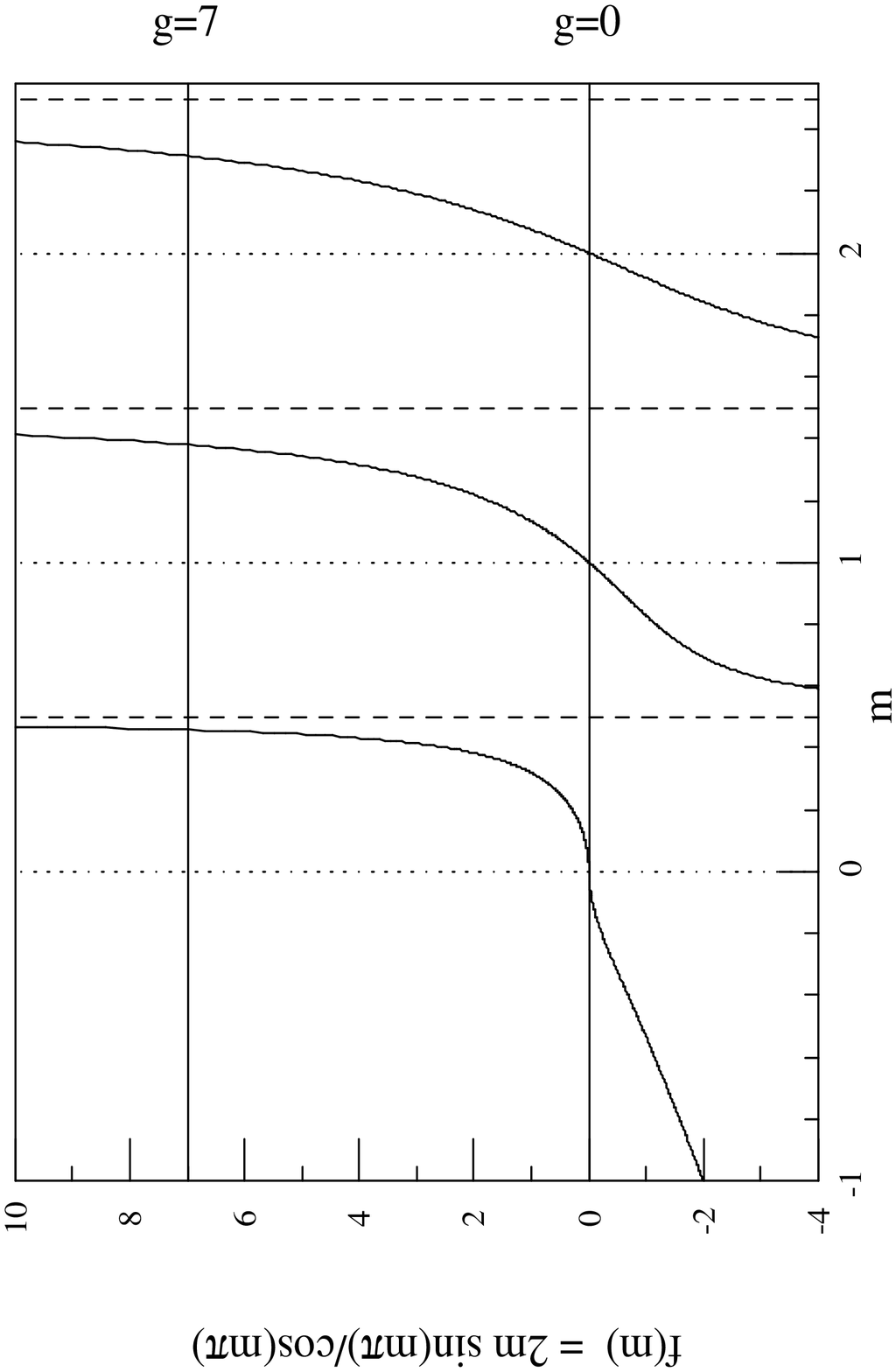,width=\linewidth}
\caption{}
\end{minipage}
\, \hfill \,
\end{figure}
\hfill 

\end{document}